**Surveys on the Existence of Extraterrestrial Intelligent Life**

**and Effects of Revealing Expert Consensus**

Omer Eldadi[1], Gershon Tenenbaum[1] and Abraham Loeb[2]

1. B. Ivcher School of Psychology, Reichman University, Herzliya, Israel

2. Department of Astronomy, Harvard University, Cambridge, MA, USA

**Corresponding Address**:

Omer Eldadi

B. Ivcher School of Psychology

Reichman University

The University 8, Herzliya, Israel

Email: Omereldadi@gmail.com

# Abstract

Vickers et al. (2025) established that 58.20% of astrobiology experts believe intelligent extraterrestrial life likely exists, providing the first empirical baseline for public comparison. We surveyed 6,114 highly educated and scientifically engaged individuals (77.60% bachelor's degree+; 67.99% high-to-very-high scientific engagement) to assess their beliefs about extraterrestrial intelligent life existence: (1) personal beliefs, (2) perceived social circle beliefs, (3) perceived expert beliefs, and (4) responses to expert consensus revelation. Results showed 95.01% believed extraterrestrial intelligent life exists, with 62.59% holding definitive rather than probable convictions. Participants exhibited massive pluralistic ignorance, a 'cosmic closet', underestimating social circle beliefs by 46.07 percentage points despite near-universal personal conviction. Participants also exhibited a novel 'conviction intensity gap': while overestimating expert belief prevalence (67.63% vs. 58.20% actual), they underestimated expert conviction strength, perceiving only 21.10% as holding definitive beliefs. Experimental revelation of actual consensus ($N = 5,106$; 83.51% passed manipulation check) produced negligible personal belief change ($d = -0.11$) and small social belief change ($d = 0.14$). These findings demonstrate that consensus misperception operates along two dimensions, *prevalence* and *intensity*, and that even scientifically engaged audiences resist belief revision via expert consensus information.

The scientific community's beliefs regarding extraterrestrial intelligent life remained largely undocumented through systematic empirical research until Vickers et al.'s survey[1]. The survey measured astrobiology experts' ($N = 521$) agreement with three distinct propositions: (1) the existence of basic extraterrestrial life, (2) extraterrestrial complex life, and (3) extraterrestrial intelligent life. Results revealed a graduated consensus pattern: 86.60% of astrobiologists agreed basic extraterrestrial life likely exists, 67.40% agreed regarding complex life, and 58.20% agreed that intelligent life likely exists somewhere in the Universe. These findings carry theoretical and practical significance. First, they establish that a clear majority of relevant scientific experts support the plausibility of intelligent extraterrestrial life existence. Second, the confidence drop from complex to intelligent life (9.2 percentage points) is less pronounced than the drop from basic to complex life (19.2 percentage points). This suggests that once complex life achieves scientific plausibility, the emergence of intelligence represents a comparatively smaller conceptual leap. Most critically for our purposes, this documented expert consensus provides the essential empirical baseline against which different audiences' perceptions can now be systematically compared.

## Public Beliefs About Extraterrestrial Intelligence Existence

Research on public attitudes reveals substantial belief in extraterrestrial intelligent life across diverse populations and contexts. The Glocalities study[2] surveying 26,492 respondents across 24 countries, found that 47% believe "the existence of intelligent alien civilizations in the universe". Cross-national variation was substantial, with belief rates ranging from 45% in the United States to 68% in Russia. The Pew Research Center[3] revealed that 65% of Americans "say their best guess is that intelligent life exists on other planets". These findings reveal substantial cross-national variation, with belief rates in individual countries ranging from 45% to 68%, spanning both below and above the documented expert consensus of 58.20%.

General population surveys provide limited insight into how different audiences perceive expert opinion. Studies targeting undergraduate student populations, which represent more scientifically engaged audiences, reveal dramatically higher belief rates. Persson et al.[4] surveyed 492 Swedish high school and undergraduate university students, reported that 90% believed extraterrestrial life existence. Importantly, the survey did not provide detailed definitions of what constituted "extraterrestrial life", meaning students may have been thinking about basic life, complex life, intelligent life, or some combination thereof when responding. Most students regarded the search as quite important or very important yet paradoxically did not see themselves as well informed. Critically, students who judged themselves as better informed showed higher belief rates, suggesting that engagement with scientific content may correlate with increased conviction about extraterrestrial life existence, even when that conviction exceeds expert consensus.

Extending these findings across cultural contexts, Chon-Torres et al. surveyed 1,237 students from Peruvian public and private universities[5], explicitly modeling their research on the Swedish study. Their results revealed that 92% of students "believe in the existence of life outside our planet". As with the Swedish study, the questionnaire did not specify what type of extraterrestrial life, and researchers acknowledged "it is quite likely that the student respondents were thinking about different forms of life (e.g., microbial, intelligent, etc.)" (p.362) when answering. Like the Swedish findings, Peruvian students considered themselves poorly informed (only 5% described themselves as "very well informed"), yet those rating themselves as better informed showed higher belief rates.

The documented spectrum of public beliefs (47%-92%) and expert consensus (58.20%) regarding the existence of extraterrestrial intelligent life raises fundamental questions about how individuals perceive close social circle beliefs, and expert consensus. The phenomenon of pluralistic ignorance[6], describes situations where individuals systematically misperceive peer beliefs, believing their private beliefs differ from those of their peers. This mechanism has been demonstrated across diverse domains. College students overestimate peer comfort with heavy drinking while privately feeling uncomfortable, perpetuating drinking cultures through collective misperception[7]. Applying pluralistic ignorance frameworks to scientific beliefs represents unexplored theoretical territory. Unlike social behavioral norms, which concern conventions and choices, scientific beliefs address factual propositions about the external world.

The critical research gap extends beyond merely documenting perception accuracy. Even if misperceptions exist, does revealing documented expert consensus influence personal beliefs, or does it merely correct factual knowledge about expert opinion while leaving personal convictions unchanged? If personal beliefs remain anchored despite learning about expert consensus, it suggests that convictions about extraterrestrial intelligence operate through deeper psychological mechanisms resistant to simple information provision.

**The Present Research**

This study addresses these theoretical and empirical gaps by examining two interrelated questions. First, do publics accurately perceive expert consensus on extraterrestrial intelligence? Given that undergraduate student populations hold beliefs exceeding expert consensus by 30-35 percentage points, do they recognize this discrepancy, or do they project their own elevated beliefs onto expert communities? Moreover, do misperceptions operate along multiple dimensions, not just the proportion of experts who believe (prevalence), but also how strongly those experts hold their beliefs (conviction intensity)? Second, does revealing actual expert consensus influence personal beliefs or perceptions of close social circle beliefs? Do individuals adjust their positions toward expert consensus, or do beliefs remain stable despite new information? In addition, we asked participants to evaluate several psychological components which we believe are related to extraterrestrial beliefs. These were anthropocentrism, curiosity, comfortable with ambiguity, skepticism, existential anxiety, institutional trust, science-fiction consumption, UFO/UAP (Unidentified Flying Objects or Unidentified Anomalous Phenomena) content exposure, and social media use.

## Results

### Participants

Participants' ($N = 6,114$) average age was 49.22 years ($SD = 13.96$), comprised 4,768 males (77.98%), 1,308 females (21.40%), 35 non-binary/third-gender individuals (0.57%) and three 'prefer not to say' participants (0.05%). Geographic distribution spanned 112 countries, with largest contingents from United States (38.47%), United Kingdom (6.97%), Spain (6.54%), and Canada (4.45%). The sample exhibited exceptionally high educational attainment: 40.80% with graduate or professional degrees (e.g., MA, MS, MBA, PhD, JD, MD), 36.80% with bachelor's degrees, 18.70% with secondary education or lower and 3.70% "prefer not to say". The participants' scientific engagement score on a 5-likert scale resulted in 23.26% ($N = 1,422$) rate '*Very high*', 44.73% ($N = 2,735$) '*High*', 27.78% '*Moderate*' ($N = 1,699$), 3.58% '*Low*' ($N = 219$) and 0.64% '*Not at all*' ($N = 39$). This exceptionally educated and scientifically engaged sample (77.60% bachelor's degree+; 67.99% high/very high scientific engagement) represents an ideal audience for examining consensus perception and responses to expert opinion.

**Extraterrestrial Intelligent Life Existence by Scientific Engagement Level**

One-way ANOVA examining personal belief in extraterrestrial intelligent life existence (1 - *Definitely does not exist* to 5 - *Definitely exists*) by scientific engagement level ($k = 5$) revealed a significant main effect, $F(4, 6109) = 14.81$, $p < .001$ (see Figure 1). Scheffé post-hoc comparisons ($p < .05$) indicated a consistent positive association between scientific engagement and belief strength: Level 5 (highest engagement) scored significantly higher than Levels 2 ($p < .001$), 3 ($p < .001$), and *4* ($p = .012$); Level 4 scored significantly higher than Levels 2 ($p = .021$) and 3 ($p = .004$). No significant differences emerged between the lowest engagement levels (1–3). Post-hoc comparisons involving Level 1 did not reach significance despite larger mean differences, reflecting insufficient statistical power due to the small cell size ($N = 39$) rather than absence of true effects.

**Figure 1**

*Participant Self-Belief by Scientific Engagement*

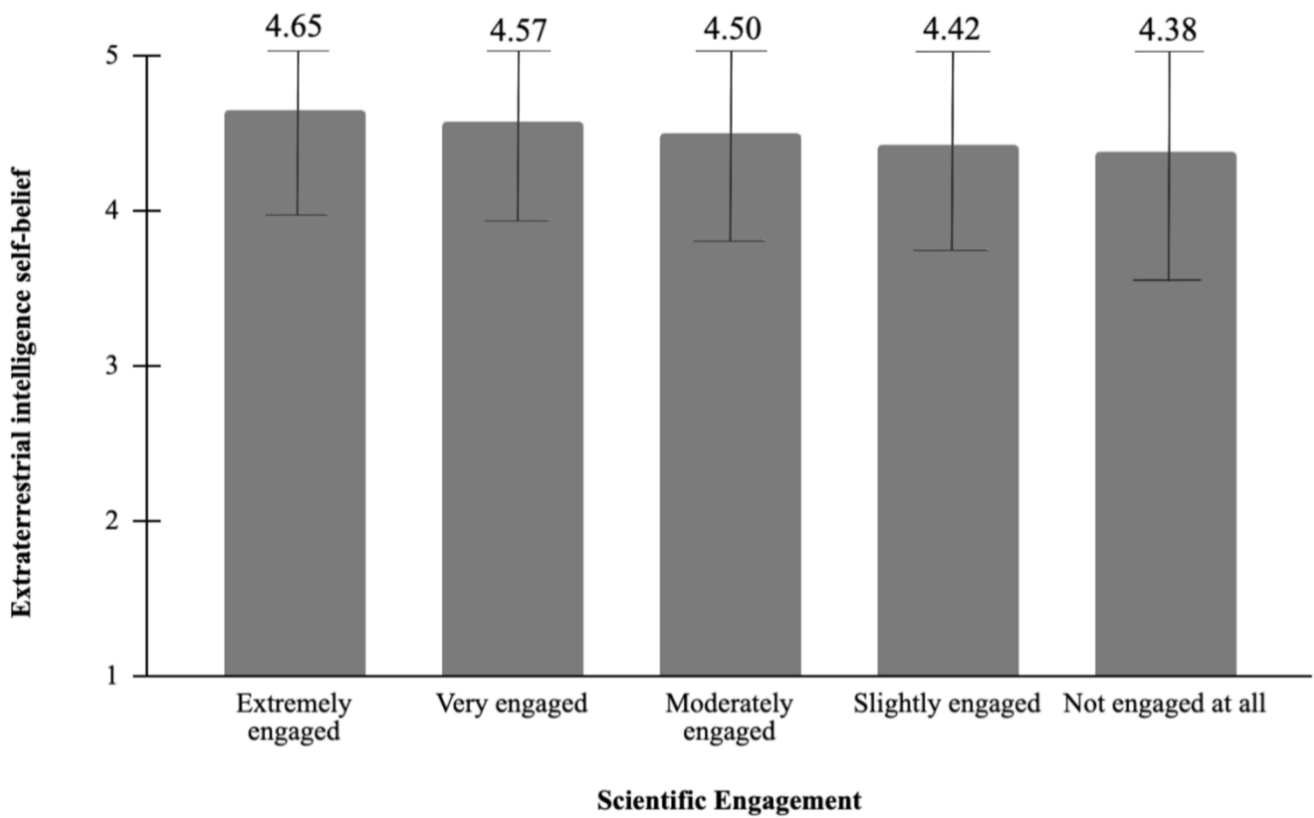

*Note.* Extremely engaged: $M = 4.65$, $SD = 0.59$, $N = 1,422$; Very engaged: $M = 4.57$, $SD = 0.61$, $N = 2,735$; Moderately engaged: $M = 4.50$, $SD = 0.66$, $N = 1,699$; Slightly engaged: $M = 4.42$, $SD = 0.71$, $N = 219$; Not engaged at all: $M = 4.38$, $SD = 1.07$, $N = 39$.

**Extraterrestrial Intelligence Existence Beliefs**

*Self-Belief:* Personal belief in extraterrestrial intelligence existence showed extreme positive skew (see Table 1). Among the participants ($N = 6,114$), 95.01% endorsed belief (ratings 4-5), comprising 62.59% '*Definitely exists*' and 32.42% '*Probably exists*'. Only 1.02% expressed skepticism (ratings 1-2) and 3.97% uncertainty (rating 3).

*Perceived Close Social Circle Beliefs:* Despite near-universal personal belief (95.01%), participants dramatically underestimated agreement within their close social circles, perceiving only 48.94% as believers, a 46.07 percentage point discrepancy (see Table 1).

*Perceived Experts Beliefs:* Participants perceived experts as convinced, estimating 67.63% endorsement, a 9.43 percentage point overestimation compared to actual consensus. However, while overestimating prevalence, participants underestimated conviction intensity. Only 21.10% attributed '*Definitely exists*' belief to experts, despite 62.59% holding such conviction themselves (self-belief, see Table 1), revealing a previously undocumented "intensity gap" in consensus perception.

**Table 1**

*Extraterrestrial Intelligence Self, Social, and Expert Beliefs Distribution*

| Response | Self-Belief | | Perceived Social Circle Beliefs | | Perceived Experts Beliefs | |
|---|---|---|---|---|---|---|
| | *N* | **% of sample** | *N* | **% of sample** | *N* | **% of sample** |
| 1 - Definitely does not exist | 12 | .20% | 54 | .88% | 44 | .72% |
| 2 - Probably does not exist | 50 | .82% | 724 | 11.84% | 446 | 7.29% |
| 3 - Uncertain/Don't know | 243 | 3.97% | 2,344 | 38.34% | 1,489 | 24.35% |
| 4 - Probably exists | 1,982 | 32.42% | 2,550 | 41.71% | 2,845 | 46.53% |
| 5 - Definitely exists | 3,827 | 62.59% | 442 | 7.23% | 1,290 | 21.10% |
| **Total "No" (1-2)** | 62 | 1.02% | 778 | 12.72% | 490 | 8.01% |
| **Total "Yes" (4-5)** | 5,809 | 95.01% | 2,992 | 48.94% | 4,135 | 67.63% |

**Psychological Predictors of Extraterrestrial Intelligence Self-Belief**

Stepwise multiple regression was performed to test nine psychological predictors of personal belief in extraterrestrial intelligent life existence ($N = 6,114$). The final model (see

Figure 2A) accounted for 13.6% of variance ($R = .37$, $R^2 = .136$, $F(6, 6107) = 160.06$, $p < .001$). UFO/UAP content exposure emerged as the strongest predictor ($\beta = .213$, $p < .001$), followed by anthropocentrism ($\beta = -.162$, $p < .001$), curiosity ($\beta = .109$, $p < .001$), institutional trust ($\beta = -.107$, $p < .001$), skepticism ($\beta = -.089$, $p < .001$), and anxiety ($\beta = .038$, $p = .001$).

**Psychological Predictors of Extraterrestrial Intelligence Social-Belief**

Stepwise multiple regression examined nine psychological predictors of perceived social circle beliefs in extraterrestrial intelligent life existence ($N = 6,114$). The final model (see Figure 2B) accounted for 1.2% of the variance ($R = .11$, $R^2 = .012$, $F(5, 6108) = 14.81$, $p < .001$). UAP content exposure emerged as the strongest predictor ($\beta = .059$, $p < .001$), followed by curiosity ($\beta = .057$, $p < .001$), social media consumption ($\beta = .035$, $p = .008$), institutional trust ($\beta = .035$, $p = .008$), and science-fiction content consumption ($\beta = .026$, $p = .049$).

**Psychological Predictors of Extraterrestrial Intelligence Experts-Belief**

Stepwise multiple regression was performed to estimate the effect of nine psychological predictors of perceived expert beliefs in extraterrestrial intelligent life existence ($N = 6,114$). The final model (see Figure 2C) accounted for 1.4% of the variance ($R = .12$, $R^2 = .014$, $F(4, 6109) = 21.80$, $p < .001$). Institutional trust emerged as the strongest predictor ($\beta = .093$, $p < .001$), followed by UAP content exposure ($\beta = .056$, $p < .001$), anxiety ($\beta = .049$, $p < .001$), and sci-fi content consumption ($\beta = .033$, $p = .011$).

**Figure 2**

*Psychological Predictors of Beliefs About Extraterrestrial Intelligence: Self, Social Circle, and Expert Consensus Perceptions*

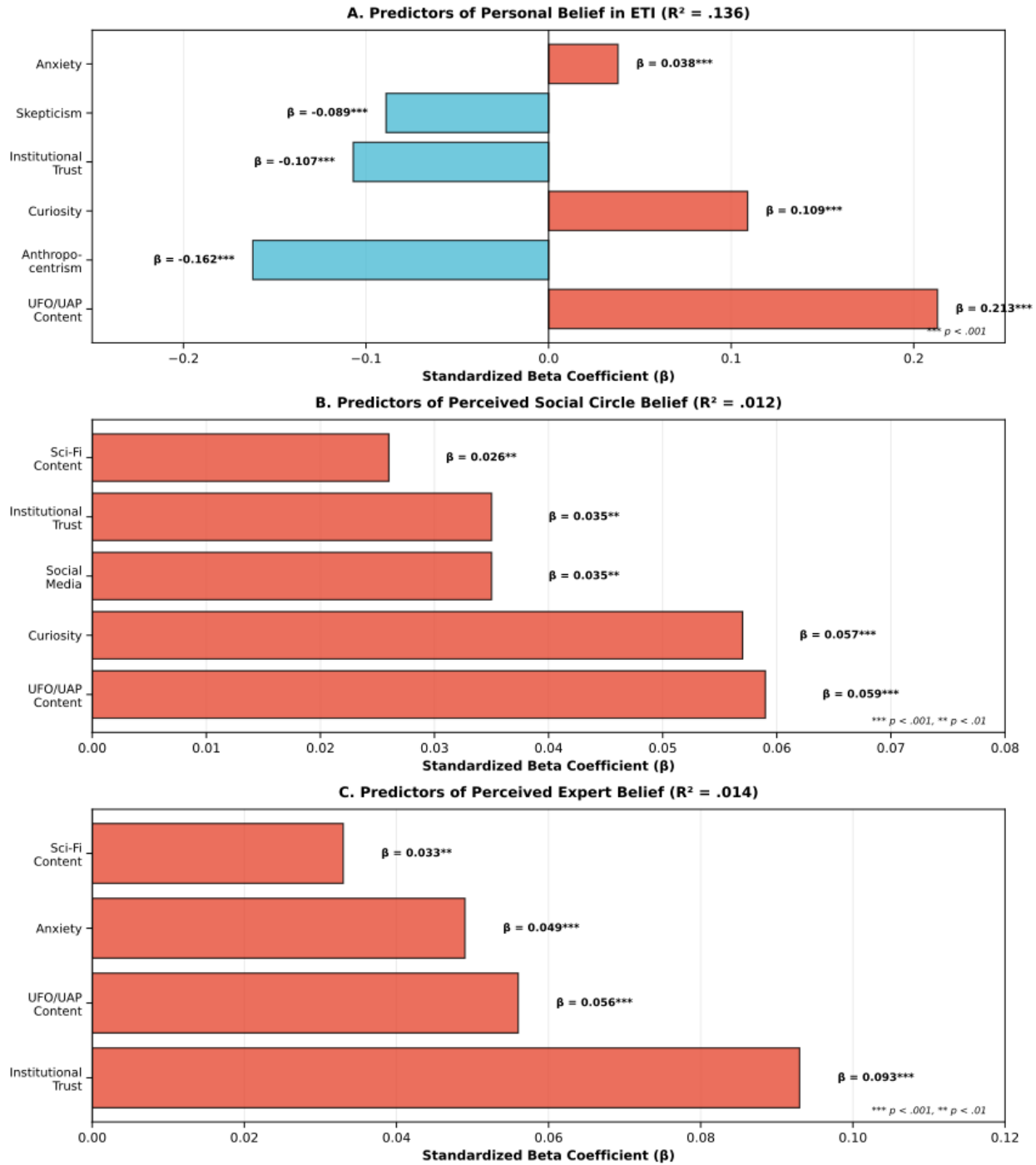

*Note.* Stepwise multiple regression with forward entry ($p < .05$ inclusion criterion). $N = 6,114$ for all models. Only significant predictors retained in final models are shown.

## Correlations Between Extraterrestrial Intelligence Belief Measures

We examined the relationships among three types of beliefs about extraterrestrial intelligence: (1) personal beliefs, (2) perceived social circle beliefs, and (3) perceived expert beliefs. Pearson correlation analyses revealed significant positive relationships among all three variables. Personal beliefs were significantly correlated with both social extraterrestrial intelligence belief (see Figure 3), $r(6112) = .22$, $p < .001$, and perceived expert beliefs, $r(6112) = .21$, $p < .001$. Perceived social circle beliefs were also significantly correlated with perceived expert beliefs, $r(6112) = .24$, $p < .001$. These results indicate weak to moderate positive associations among the three belief types, suggesting that individuals' personal beliefs about extraterrestrial intelligence are somewhat related to their perceptions of both societal and expert beliefs.

**Figure 3**

*Correlation Matrix of Extraterrestrial Intelligence Belief Measures (N = 6,114)*

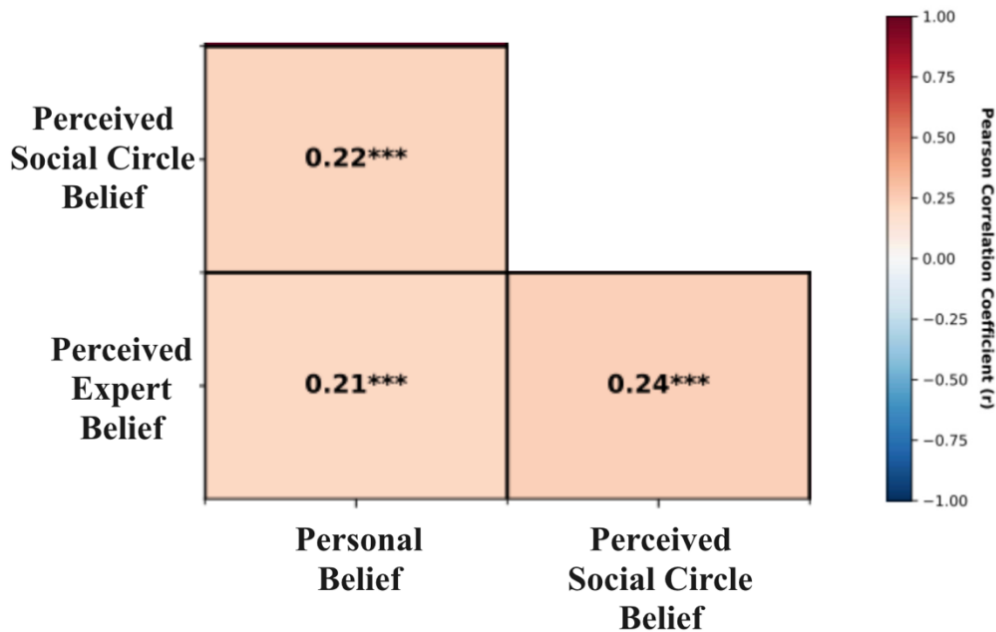

*Note.* Correlations are weak to moderate, indicating that the three belief dimensions capture distinct aspects of consensus perception. \*\*\* $p < .001$.

### The Conviction Intensity Gap

Beyond prevalence misperception, analysis revealed systematic underestimation of expert conviction strength, a previously undocumented dimension of consensus misperception. Among the 58.20% of experts who believe extraterrestrial intelligent life exists, Vickers et al. (2025) show in their Figure 1 that '*strongly agree*' responses were less than '*agree*' responses, though they did not report the exact proportions. In this study, among participants who attributed belief to experts (rating 4-5, $N = 4{,}135$): $N = 1{,}290$ (21.10%) rated '*Definitely*' (perceived as strong conviction), and $N = 2{,}845$ (46.53%) rated '*Probably*' (perceived as moderate conviction). This reveals an intensity gap: participants projected predominantly moderate conviction onto experts (see Figure 4) while participants themselves held definitive beliefs (62.59% rated '*Definitely*'), yet assumed experts remained tentatively convinced despite similar belief prevalence.

**Figure 4**

*Personal Beliefs, Social Perceptions, and Expert Consensus: Actual vs. Perceived*

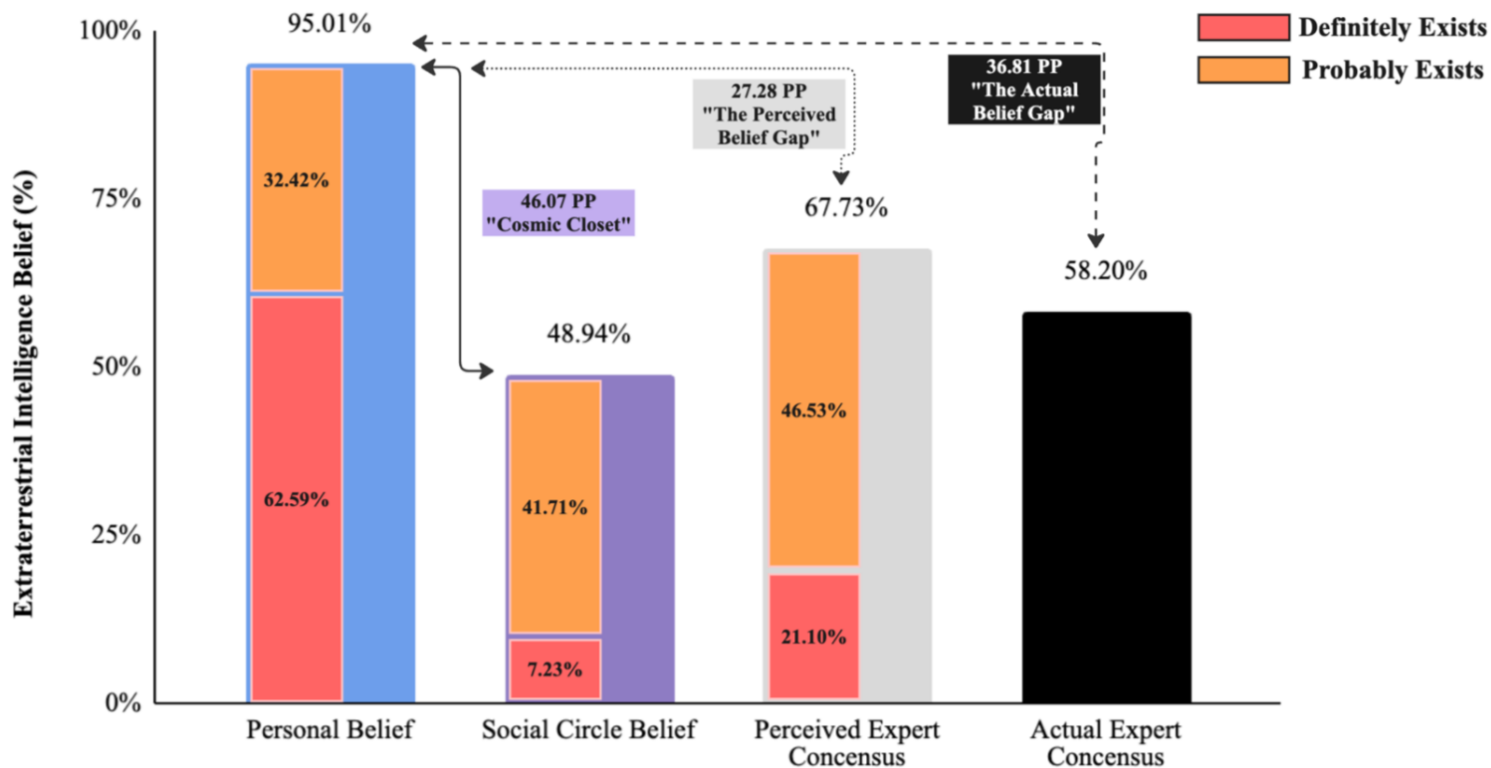

**Intervention Effect: Revealing Expert Consensus**

*Manipulation Check*

Of the 6,114 participants, 5,106 (83.51%) correctly identified the 58.20% expert consensus on the manipulation check, forming the intervention analysis sample. Participants who passed showed identical baseline patterns to the full sample, confirming that intervention effects were not driven by inattention or differential attrition.

**Post-Intervention Changes in Personal Beliefs**

We examined pre-post results of participants who answered correctly on the manipulation check ($N = 5,106$). The intervention produced minimal change in personal beliefs. A repeated measures ANOVA revealed a significant main effect on personal beliefs, $F(1, 5101) = 27.57$, $p < .001$. Pre-intervention belief ($M = 4.53$, $SD = .63$) decreased to post-intervention ($M = 4.46$, $SD = .64$), representing a statistically significant but practically negligible change ($d = -0.11$, see Figure 5). The interaction between pre-post measurements and degree of scientific engagement was not significant, $F(4, 5101) = 1.40$, $p = .23$, indicating the lack of intervention effect across engagement levels. While overall belief prevalence barely changed (94.71%), the proportion holding '*Definite exists*' beliefs decreased from 60.54% to 55.66% (−4.88pp), with corresponding increases in '*Probably exists*' beliefs from 34.33% to 39.05% (+4.72pp).

**Post-Intervention Changes in Close Social Circle Beliefs**

Perceived social circle beliefs showed a significant positive shift. A repeated measures ANOVA revealed a significant main effect of $F(1, 5101) = 28.48$, $p < .001$. Presented in Figure 5, pre-intervention perceived social belief ($M = 3.37$, $SD = .81$) increased to post-intervention ($M = 3.48$, $SD = .77$, $d = 0.14$). The interaction between pre-post measurements and degree of scientific engagement was not significant, $F(4, 5101) = 1.01$, $p = .403$, indicating the intervention effect was consistent across engagement levels. Perceived social circle beliefs increased from 48.22% to 53.29% (+5.07pp), partially correcting pluralistic ignorance but leaving a 41.42 percentage point residual gap post intervention between personal (94.71%) and perceived social beliefs (53.29%).

**Figure 5**

*Pre-Post Extraterrestrial Intelligence Personal and Perceived Social Circle Beliefs*

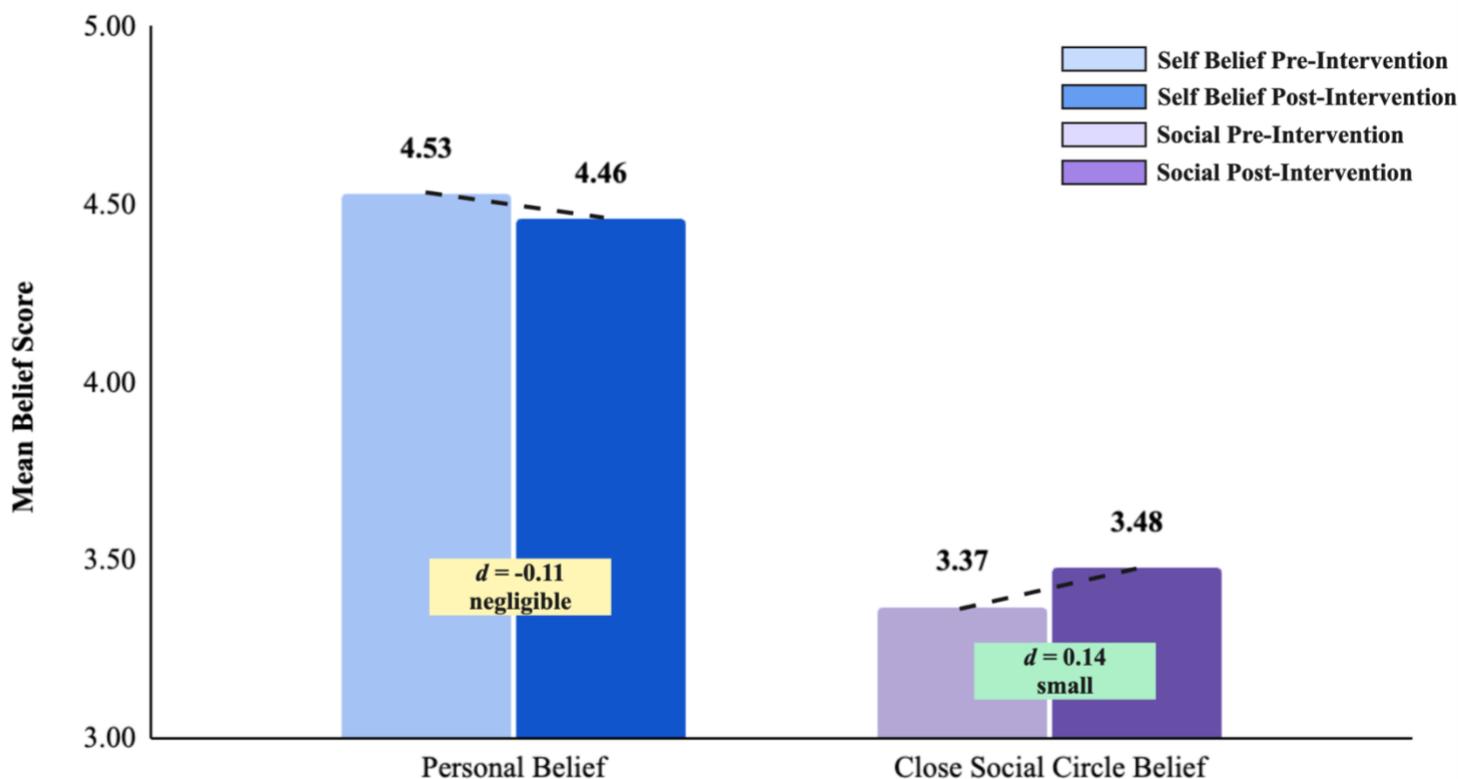

## Discussion

Our investigation of 6,114 highly educated and scientifically engaged individuals reveals belief dynamics that fundamentally challenge conventional science communication paradigms. Four principal findings emerge: (1) near-universal belief in extraterrestrial intelligence (95.01%) far exceeding actual expert consensus (58.20%), (2) massive pluralistic ignorance with a 46.07 percentage point gap between personal and perceived close social beliefs, (3) the discovery of a previously undocumented "conviction intensity gap" in expert perception, and (4) revealing actual expert consensus produced negligible effects on personal beliefs ($d = -0.11$), demonstrating profound resistance to informational interventions even among audiences maximally primed for scientific information processing.

The phenomenon we term the "cosmic closet" represents one of the largest documented cases of pluralistic ignorance in science communication literature. While 95.01% of participants privately believe extraterrestrial intelligence exists, they estimate only 48.94% of their social circles share this view. This misperception creates a self-perpetuating cycle: individuals remain silent about their beliefs, interpreting others' silence as skepticism, which reinforces their own reticence. Beliefs about extraterrestrial intelligence existence remain socially constrained despite near-universal private acceptance. This silence likely stems from multiple psychological and social factors. Concerns about appearing unscientific, particularly among educated populations, may suppress expression[8,9,10]. The historical association of extraterrestrial intelligence beliefs with fringe communities and conspiracy theories creates reputational risk. Additionally, the absence of established social scripts for discussing cosmic questions leaves individuals uncertain about appropriate discourse norms. Our intervention's partial success, increasing perceived social beliefs by 5.07 percentage points, suggests consensus information can begin recalibrating social perceptions, though the residual 41.42 percentage point gap indicates single-exposure messaging remains insufficient.

Our discovery of the conviction intensity gap reveals that consensus misperception operates along two independent dimensions: prevalence and intensity. While participants overestimated the proportion of believing experts (67.63% vs. 58.20% actual), they simultaneously underestimated conviction strength among those believing experts, with only 21.10% attributing 'Definitely exists' beliefs to experts. This asymmetry, where participants attribute definitive beliefs to themselves (62.59%) but only tentative conviction to experts, suggests a fundamental attribution error in consensus perception[11]. Individuals may view their own beliefs as products of careful reasoning while assuming experts' beliefs remain probabilistic and evidence-contingent. This finding has profound implications for science communication. Current strategies typically report simple prevalence statistics ("X% of scientists believe...")[12,13], neglecting the conviction dimension. Our results suggest that communicating both dimensions: how many experts believe and how strongly they believe, may more accurately convey scientific consensus and potentially enhance its influence on public opinion.

The psychological predictors revealed distinct patterns across belief dimensions. UFO/UAP content exposure emerged as the strongest predictor of personal beliefs ($\beta = .213$), though the cross-sectional design precludes causal inference, whether exposure cultivates belief or believers seek confirming content remains undetermined. Anthropocentrism showed a strong negative association ($\beta = -.162$), consistent with theoretical frameworks linking cosmic perspective to reduced human exceptionalism. Lower institutional trust predicted higher personal belief ($\beta = -.107$), suggesting that distrust in official narratives may paradoxically increase openness to extraterrestrial possibilities. Notably, the variance explained in perceived social beliefs (1.2%) and expert beliefs (1.4%) was markedly lower than for personal beliefs (13.6%), indicating that social perceptions operate through different mechanisms, potentially heuristic processes or social projection rather than deliberative assessment of others' positions.

The subtle finding that "definite" beliefs decreased by 4.88 percentage points while "probable" beliefs correspondingly increased suggests the intervention produced epistemic humility rather than belief revision. This modulation of certainty without abandoning core beliefs indicates consensus information may influence metacognitive confidence even when failing to change belief direction. Participants adjusted the strength of their conviction while maintaining their fundamental position, a pattern consistent with motivated reasoning frameworks where individuals assimilate challenging information in ways that preserve core beliefs.

## Conclusion

This study establishes that beliefs about extraterrestrial intelligence among scientifically engaged publics operate through mechanisms distinct from conventional scientific beliefs, carrying three key implications. First, the 36.81 percentage point gap between public belief (95.01%) and expert consensus (58.20%) among highly educated audiences suggests not scientific illiteracy but rather different epistemological frameworks for evaluating cosmic questions.

Second, the discovery of massive pluralistic ignorance, the "cosmic closet", indicates the primary barrier to public discourse is not skepticism but misperceived social norms. With 95.01% privately believing but assuming majority skepticism, individuals self-censor, perpetuating false perceptions of minority status. This finding suggests communication strategies should prioritize revealing existing consensus among peers rather than attempting to change beliefs already near ceiling.

Third, the conviction intensity gap represents a novel contribution to understanding consensus perception, demonstrating that people track not just what experts believe but how strongly they believe it. This two-dimensional model of consensus perception: prevalence and intensity, opens new theoretical and practical avenues for science communication research beyond the extraterrestrial domain.

Our findings suggest the public, at least among scientifically engaged communities, exhibits greater readiness for potential discovery than experts assume. The challenge lies not in convincing skeptics but in creating social conditions where existing believers feel comfortable expressing their convictions, enabling informed public discourse about humanity's possible cosmic companions.

## Method

### Participants

Participants were recruited via social media platforms such as Medium, Twitter, LinkedIn and WhatsApp. Data collection occurred between November 10-16, 2025, through an embedded survey link. To be eligible for the study, participants were at least 18 years old. Sample size was calculated using G*Power 3[14] for the primary analyses. To ensure the study was sufficiently powered to detect even very small effects, parameters were set at an effect size of $f = 0.1$ with 80% power at $\alpha = .01$ for the primary analyses. This calculation yielded a required sample size of approximately $N = 1,800$, which our achieved sample exceeds. A total of $N = 6,114$ adults completed the survey.

For intervention analyses examining consensus information effects, we additionally required correct responses to a manipulation check verifying participants read and processed the expert consensus information. This criterion (passing manipulation check) identified $N = 5,106$ participants (83.51% of total sample) for intervention analyses.

### Measurements (see appendix A for full questionnaire)

*Demographic Questionnaire* (DQ). The DQ includes items on country, age, gender, education level and levels of engagement with science-related topic.

*Psychological Predictors*. Beyond demographic factors, we measured psychological orientations theoretically relevant to beliefs about extraterrestrial intelligence: *anthropocentrism* (humans' specialness)*, curiosity, tolerance for uncertainty, skepticism, existential anxiety, institutional trust*, and media exposure: *science fiction* consumption, *UFO/UAP* content exposure and *social media* use.

The psychological variables were assessed via single-item self-report measures using 5-point Likert scales: (1 = 'Not at all' to 5 = 'Extremely'). Media consumption was assessed via single items for science fiction exposure (weekly hours), UFO/UAP content frequency (1 = Never to 5 = Very frequently), and social media use (daily hours). Complete items available are presented in appendix A.

*Baseline Belief Measures.* Three parallel items assess beliefs before intervention:

- Personal belief: "To what extent do you believe that intelligent extraterrestrial life exists somewhere in the universe beyond Earth?"

- Perceived social circle belief: "To what extent do you think most people in your social circle believe that intelligent extraterrestrial life exists somewhere in the universe beyond Earth?"

- Perceived expert belief: "To what extent do you think most astrobiology experts believe that intelligent extraterrestrial life exists somewhere in the universe beyond Earth?"

All items used identical 5-point Likert-type scales ranging from $1 - 5$ ( 1 - *Definitely does not exist*, 2 - *Probably does not exist*, 3 - *Uncertain/don't know*, 4 - *Probably exists*, 5 - *Definitely exists*.

***Intervention***

Participants read: "Recent research by Vickers et al. (2025) surveyed over 500 astrobiology experts about extraterrestrial life. The study found that 58.20% of experts believe intelligent extraterrestrial life likely exists somewhere in the universe. This represents the first systematic documentation of scientific expert opinion on this question".

- Manipulation Check. "According to the research mentioned, what percentage of experts believe intelligent extraterrestrial life likely exists?" Options: 58.20% (correct), other (incorrect). Incorrect responses indicated failure to process consensus information.

**Procedure**

After providing informed consent, participants completed demographic questions, psychology factors and baseline belief measures. They then read the intervention text presenting actual expert consensus regarding the likelihood of extraterrestrial intelligent life. The manipulation check followed immediately, where participants were asked to select the correct expert consensus of 58.20% from 5 options. If they chose a wrong option, they were excluded for the pre/post comparison data analysis. Finally, participants re-rated their personal and social circle beliefs. The survey concluded with debriefing information about the study's purpose. Participants re-rated personal and social circle beliefs using identical scales, allowing within-subject comparison of pre-post changes.

**Data Analysis**

All analyses were conducted using SPSS Statistics (Version 29). Statistical significance was set at $\alpha = .05$ for all tests. Descriptive analyses examined belief distributions using frequency tables and percentages. We calculated endorsement rates by collapsing 5-point scales into "Yes" (ratings 4-5), "Uncertain/don't know" (rating 3), and "No" (ratings 1-2) categories. Between-group comparisons employed one-way ANOVA examining personal beliefs across five scientific engagement levels, with Scheffé post-hoc tests for pairwise comparisons. Intervention effects were analyzed using repeated measures ANOVA with pre-post as the within-subjects factor and scientific engagement as the between-subjects factor. This design tested both main effects of the intervention and potential moderating effects of scientific engagement on belief change. Separate analyses examined personal beliefs and perceived social circle beliefs as dependent variables. Pluralistic ignorance was assessed by comparing personal beliefs (self-rated) against perceived social circle beliefs at baseline. Effect sizes were calculated as *Cohen's d* with benchmarks of $d = 0.2$ (small), $d = 0.5$ (medium), and $d = 0.8$ (large) following Cohen[15].

Psychological predictors were examined using stepwise multiple regression with forward entry. Nine predictor variables (*anthropocentrism, curiosity, tolerance for uncertainty, existential anxiety, institutional trust, science fiction consumption, UFO/UAP content exposure, social media use, and skepticism*) were entered, with inclusion criterion of $p < .05$. Separate regression models predicted (i) personal beliefs, (ii) perceived social circle beliefs, and (iii) perceived expert beliefs. Correlational analyses used Pearson product-moment correlations to examine relationships among the three belief measures (personal, perceived social circle, and perceived expert beliefs). Manipulation check success (correctly identifying 58.20% as the expert consensus figure) served as the inclusion criterion for intervention analyses, reducing the sample from $N = 6,114$ to $N = 5,106$.

**Acknowledgements**

**A.L. was supported in part by Harvard's Black Hole Initiative and the Galileo Project.**

**Author Contributions**

All authors conceived the study, designed the survey, collected data, performed analyses, and wrote the manuscript.

**Competing Interests**

The authors declare no competing interests.

**Data Availability**

Raw data available at:

https://docs.google.com/spreadsheets/d/1VNIJCzuAmKjb5LcEtS4iHjJi6OnXyZ7mwGFcTeB8V ck/edit?usp=sharing

**Ethics Statement**

This study was approved by Reichman University IRB Ethics Committee.

Ethical clearance number: P_2025220

All participants provided informed consent.

# References

[1] Vickers, P, Gardiner, E, Gillen, C, Hyde, B, Jeancolas, C, Finnigan, SM, Nováková, JN, Strandin, H, Tasdan, U, Taylor, H & McMahon, S. (2025), Surveys of the scientific community on the existence of extraterrestrial life, *Nature Astronomy,9*(1), pp. 16-18. https://doi.org/10.1038/s41550-024-02451-0

[2] Lampert, M., & Papadongonas, P. (2017). Majority of humanity say we are not alone in the universe: Values-based learnings from the Glocalities survey in 24 countries. *Glocalities*. https://glocalities.com/reports/majority-of-humanity-say-we-are-not-alone-in-the-universe

[3] Kennedy, C., & Lau, A. (2021, June 30). *Most Americans believe in intelligent life beyond Earth; few see UFOs as a major national security threat*. Pew Research Center. https://www.pewresearch.org/short-reads/2021/06/30/most-americans-believe-in-intelligent-life-beyond-earth-few-see-ufos-as-a-major-national-security-threat/

[4] Persson E., Capova K.A., Li Y. (2019). Attitudes towards the scientific search for extraterrestrial life among Swedish high school and university students. *International Journal of Astrobiology, 18*(3):280-288. doi:10.1017/S1473550417000556

[5] Chon-Torres O.A., Ramos Ramírez J.C., Hoyos Rengifo F., Choy Vessoni R.A., Sergo Laura I., Ríos-Ruiz F.G.A., Murga-Moreno C.A., Alvarado Pino J.P.L., Yance-Morales X. (2020). Attitudes and perceptions towards the scientific search for extraterrestrial life among students of public and private universities in Peru. *International Journal of Astrobiology, 19*, 360–368. https://doi.org/10.1017/S1473550420000130

[6] Miller, D. T., & McFarland, C. (1987). Pluralistic ignorance: When similarity is interpreted as dissimilarity. *Journal of Personality and Social Psychology, 53*(2), 298–305. https://doi.org/10.1037/0022-3514.53.2.298

[7] Prentice, D. A., & Miller, D. T. (1993). Pluralistic ignorance and alcohol use on campus. *Journal of Personality and Social Psychology, 64*(2), 243-256.

[8] Goffman, E. (1963). *Stigma: Notes on the management of spoiled identity*. Simon & Schuster.

[9] Major, B., & O'Brien, L. T. (2005). The social psychology of stigma. *Annual Review of Psychology, 56*, 393-421. https://doi.org/10.1146/annurev.psych.56.091103.070137

[10] Lantian, A., Muller, D., Nurra, C., & Douglas, K. M. (2018). Stigmatized beliefs: Conspiracy theories, anticipated negative evaluation of the self, and fear of social exclusion. *European Journal of Social Psychology, 48*(7), 939-954. https://doi.org/10.1002/ejsp.2498

[11] Ross, L. (1977). The intuitive psychologist and his shortcomings: Distortions in the attribution process. In L. Berkowitz (Ed.), *Advances in experimental social psychology* (Vol. 10, pp. 173-220). Academic Press. https://doi.org/10.1016/S0065-2601(08)60357-3

[12] van der Linden, S. (2021). The Gateway Belief Model (GBM): A review and research agenda for communicating the scientific consensus on climate change. *Current Opinion in Psychology*, *42*, 7–12. https://doi.org/10.1016/j.copsyc.2021.01.005

[13] van Stekelenburg, A., Schaap, G., Veling, H., van 't Riet, J., & Buijzen, M. (2022). Scientific-consensus communication about contested science: A preregistered meta-analysis. *Psychological Science, 33*(12), 1989-2008. https://doi.org/10.1177/09567976221083219

[14] Faul, F., Erdfelder, E., Lang, A. G., & Buchner, A. (2007). G*Power 3: A flexible statistical

power analysis program for the social, behavioral, and biomedical sciences. *Behavior

Research Methods, 39*(2), 175-191. https://doi.org/10.3758/BF03193146

[15] Cohen, J. (1988). Statistical power analysis for the behavioral sciences (2nd ed.). Routledge.

https://doi.org/10.4324/9780203771587

**Appendix A – Survey**

**Demographic Questions (DQ)**
**1. In which country do you currently reside?**
**2. How old are you?**
**3. What is your gender?**

| 1 | 2 | 3 | 4 |
|---|---|---|---|
| Male | Female | Non-binary | Prefer not to say |

**4. What is the highest level of education you have completed?**

| 5 | 4 | 3 | 2 | 1 |
|---|---|---|---|---|
| Graduate or professional degree (MA, MS, MBA, PhD, Law Degree, Medical Degree etc) | University bachelor's degree | Secondary | Primary | Prefer not to say |

*5.* **How would you describe your level of engagement with science-related topics?**

| 5 | 4 | 3 | 2 | 1 |
|---|---|---|---|---|
| Extremely engaged | very engaged | Moderately engaged | Slightly engaged | Not engaged at all |

**6. To what extent do you see yourself as curious and novelty-seeking?**

| 5 | 4 | 3 | 2 | 1 |
|---|---|---|---|---|
| Extremely | Considerably | Moderately | Slightly | Not at all |

**7. To what extent do you feel comfortable with ambiguity?**

| 5 | 4 | 3 | 2 | 1 |
|---|---|---|---|---|
| Extremely | Considerably | Moderately | Slightly | Not at all |

**8. To what extent do you see yourself as skeptical?**

| 5 | 4 | 3 | 2 | 1 |
|---|---|---|---|---|
| Extremely | Considerably | Moderately | Slightly | Not at all |

**9. To what extent do you think humans are special compared to other potential intelligent life in the universe (anthropocentrism)?**

| 5 | 4 | 3 | 2 | 1 |
|---|---|---|---|---|
| Extremely | Considerably | Moderately | Slightly | Not at all |

**10. To what extent do you trust institutions (government, scientific) to tell the truth about significant discoveries?**

| 5 | 4 | 3 | 2 | 1 |
|---|---|---|---|---|
| Extremely | Considerably | Moderately | Slightly | Not at all |

**11. To what extent do you consume science fiction content (movies, TV shows, books)?**

| 5 | 4 | 3 | 2 | 1 |
|---|---|---|---|---|
| Over 5 hours weekly | 3-5 hours weekly | 1-2 hours weekly | Less than 1 hour weekly | None |

**12. To what extent do you spend on social media platforms daily?**

| 5 | 4 | 3 | 2 | 1 |
|---|---|---|---|---|
| More than 6 hours a day | 4-6 hours a day | 2-4 hours a day | 1-2 hours a day | 0-1 hours a day |

**13. To what extent do you seek out or encounter content about UFOs/UAPs?**

| 5 | 4 | 3 | 2 | 1 |
|---|---|---|---|---|
| Very frequently | Often | Occasionally | Rarely | Never |

**14. To what extent does thinking about the unknown in the universe cause you anxiety?**

| 5 | 4 | 3 | 2 | 1 |
|---|---|---|---|---|
| Extremely | Considerably | Moderately | Slightly | Not at all |

**Baseline Measurements**
**15. To what extent do you believe that intelligent extraterrestrial life exists somewhere in the universe beyond Earth?**

| 5 | 4 | 3 | 2 | 1 |
|---|---|---|---|---|
| Definitely exists | Probably exists | Uncertain/Don't know | Probably does not exist | Definitely does not exist |

**16. To what extent do you think most people in your social circle (friends, family, colleagues) believe that intelligent extraterrestrial life exists somewhere in the universe beyond Earth?**

| 5 | 4 | 3 | 2 | 1 |
|---|---|---|---|---|
| Definitely exists | Probably exists | Uncertain/Don't know | Probably does not exist | Definitely does not exist |

**17. To what extent do you think most astrobiology experts believe that intelligent extraterrestrial life exists somewhere in the universe beyond Earth?**

| 5 | 4 | 3 | 2 | 1 |
|---|---|---|---|---|
| Definitely exists | Probably exists | Uncertain/Don't know | Probably does not exist | Definitely does not exist |

**Intervention Text (and manipulation check):**

**18. A recent academic paper of astrobiology experts published in *Nature Astronomy* by Vickers et al., 2025 found that: 58.20% of astrobiology experts believe intelligent extraterrestrial life likely exists.**
**These findings represent the current consensus among scientists specializing in the search for life beyond Earth.**
**Based on the information provided, what percentage of astrobiology experts believe intelligent extraterrestrial life likely exists?**

| 5 | 4 | 3 | 2 | 1 |
|---|---|---|---|---|
| Under 30% | 30% | 58.20% | 80% | 100% |

**19. After reading the information above, to what extent do you now believe that intelligent extraterrestrial life exists somewhere in the universe beyond Earth?**

| 5 | 4 | 3 | 2 | 1 |
|---|---|---|---|---|
| Definitely exists | Probably exists | Uncertain/Don't know | Probably does not exist | Definitely does not exist |

**20. After reading the information above, to what extent do you think most people in your social circle believe that intelligent extraterrestrial life exists somewhere in the universe beyond Earth?**

| 5 | 4 | 3 | 2 | 1 |
|---|---|---|---|---|
| Definitely exists | Probably exists | Uncertain/Don't know | Probably does not exist | Definitely does not exist |

---

### Thank you for completing this survey!

This research examines how people's beliefs about extraterrestrial intelligence compare to expert scientific opinion, and whether learning about expert consensus changes those beliefs. Understanding these patterns can help improve science communication about astrobiology and the search for extraterrestrial intelligence.

Questions or concerns?
Contact: omer.eldadi@post.runi.ac.il
Want to learn more about the research cited in this survey?
Vickers et al. (2025): https://doi.org/10.1038/s41550-024-02451-0
Thank you again for your valuable contribution to this research!